\documentclass[a4paper, conference]{IEEEtran}
\usepackage{cite}
\usepackage{amsmath,amssymb,amsfonts}
\usepackage{algorithmic}
\usepackage{textcomp}
\usepackage{url}
\usepackage{graphicx}
\usepackage{xcolor}
\usepackage[inline]{enumitem}
\usepackage{cite}
\usepackage{tikz}
\usepackage{pgfplots}
\pgfplotsset{compat=newest}
\pgfplotsset{plot coordinates/math parser=false} 

\usepackage{scalerel}
\usetikzlibrary{svg.path}
\definecolor{orcidlogocol}{HTML}{A6CE39}
\tikzset{
  orcidlogo/.pic={
    \fill[orcidlogocol] svg{M256,128c0,70.7-57.3,128-128,128C57.3,256,0,198.7,0,128C0,57.3,57.3,0,128,0C198.7,0,256,57.3,256,128z};
    \fill[white] svg{M86.3,186.2H70.9V79.1h15.4v48.4V186.2z}
                 svg{M108.9,79.1h41.6c39.6,0,57,28.3,57,53.6c0,27.5-21.5,53.6-56.8,53.6h-41.8V79.1z M124.3,172.4h24.5c34.9,0,42.9-26.5,42.9-39.7c0-21.5-13.7-39.7-43.7-39.7h-23.7V172.4z}
                 svg{M88.7,56.8c0,5.5-4.5,10.1-10.1,10.1c-5.6,0-10.1-4.6-10.1-10.1c0-5.6,4.5-10.1,10.1-10.1C84.2,46.7,88.7,51.3,88.7,56.8z};
  }
}
\newcommand\orcidicon[1]{\href{https://orcid.org/#1}{\mbox{\scalerel*{
\begin{tikzpicture}[yscale=-1,transform shape]
\pic{orcidlogo};
\end{tikzpicture}
}{|}}}}
\usepackage{hyperref} 

\newcommand{\sS}[0]{\mathcal{S}}
\newcommand{\sE}[0]{\mathcal{E}}
\newcommand{\sQ}[0]{\mathcal{Q}}
\newcommand{\sR}[0]{\mathcal{R}}
\newcommand{\sU}[0]{\mathcal{U}}

\newcommand{\SNn}[0]{\mathrm{SN}_0}

\DeclareMathAlphabet{\mathpzc}{OT1}{pzc}{m}{n}

\def\arXivPrint{1}  

\makeatletter
\newcommand\gobblestar
    {\futurelet\@let@token\@dogobblestar}

\def\@dogobblestar
    {\let\next=\relax
     \ifx\@let@token*%
        \def\next{\@gobble}%
     \else\ifx\@let@token[%
         \def\next{\@gobbleoptional}%
     \fi\fi
     \next}

\def\@gobbleoptional[#1]{}
\makeatother

\newcommand\doSingleLine[1]
    {\let\normalnewline=\\
     \let\\=\gobblestar
     #1%
     \let\\=\normalnewline}
\begin{document}

\title{Radar Resource Management for Active Tracking Using Split-Aperture Phased Arrays%
}

\author{\IEEEauthorblockN{Pepijn B. Cox \orcidicon{0000-0002-8220-7050} and Wim L. van Rossum \orcidicon{0000-0003-2618-164X}}
\IEEEauthorblockA{\textit{Radar Technology Department, TNO} \\
The Hague, The Netherlands \\
\{pepijn.cox, wim.vanrossum\}@tno.nl}
}

%
%
\ifx\arXivPrint\undefined\else

\makeatletter
\twocolumn[{
\vspace{2cm}
This paper has been accepted for publication at the

\vspace{1cm}
\centerline{\textbf{\huge{2024 IEEE Radar Conference}}}

\vspace{5cm}

\vspace{1cm}
\textbf{Citation}\\
P.B. Cox and W.L. van Rossum, ``\doSingleLine{\@title},'' in \textit{Proceedings of the 2024 IEEE Radar Conference}, pp --, Denver, Collorado, USA, May 2024.

\vspace{1cm}

\definecolor{commentcolor}{gray}{0.9}
\newcommand{\commentbox}[1] {\colorbox{commentcolor}{\parbox{\linewidth}{#1}}}

\vspace{4cm}
\commentbox{
	\vspace*{0.2cm}
	\hspace*{0.2cm}More papers from P.B. Cox can be found at\\~\\
	\centerline{\large{\url{https://orcid.org/0000-0002-8220-7050}}}\\~\\
	\hspace*{0.2cm}and of W.L. van Rossum at\\
	\centerline{\large{\url{https://scholar.google.com/citations?user=Lh1u0qMAAAAJ}}}\\~\\
	\vspace*{0.2cm}
}

\vspace{3cm}
\textcopyright 2024 IEEE. Personal use of this material is permitted. Permission from IEEE must be obtained for all other uses, in any current or future media, including reprinting/republishing this material for advertising or promotional purposes, creating new collective works, for resale or redistribution to servers or lists, or reuse of any copyrighted component of this work in other works.
}]
\clearpage
\makeatother

\fi
%
%

\maketitle

\begin{abstract}
Flexible front-end technology will become available in future multifunction radar systems to improve adaptability to the operational theatre. A potential concept to utilize this flexibility is to subdivide radar tasks spatially over the array, the so-called \emph{split-aperture phased array} (SAPA) concept. As radars are generally designed for their worst-case scenario, e.g., small targets at a large range, the power-aperture budget can be excessive for targets that do not fall within that class. To increase efficiency of the time budget of the radar front-end, the SAPA concept could be applied. In this paper, the SAPA concept is explored to assign radar resources for active tracking tasks of many targets. To do so, we formulate and solve the radar resource management problem for the SAPA concept by employing the \emph{quality of service based resource allocation model} (Q-RAM) framework. It will be demonstrated by a simulation example that a radar can maintain a larger numbers of active tracking tasks when using the SAPA concept compared to the case that only the full array can be used per task.
\end{abstract}


\section{Introduction}

\begin{figure}[!b]%
\vspace{-0.5cm}
\centering%
\definecolor{mycolor1}{rgb}{0.00000,0.44706,0.74118}%
\definecolor{mycolor2}{rgb}{0.85098,0.32549,0.09804}%
\definecolor{mycolor3}{rgb}{0.92900,0.69400,0.12500}%
\begin{tikzpicture}[font=\footnotesize]
	\draw [draw=black, thick] (0,0) rectangle ++(7,2.5);
	\node at (3.5,2.5) [above=1] {Phased array};
	\draw [draw=mycolor1, fill=mycolor1!40!white] (0.1,0.1) rectangle ++(2.10,2.3) node[pos=.5] {{\color{mycolor1!80!black}Task 1}};
	\draw [draw=mycolor2, fill=mycolor2!40!white] (2.30,0.1) rectangle ++(3.05,2.3) node[pos=.5] {{\color{mycolor2!80!black}Task 2}};
	\draw [draw=mycolor3, fill=mycolor3!40!white] (5.45,0.1) rectangle ++(1.45,2.3) node[pos=.5] {{\color{mycolor3!80!black}Task 3}};
\end{tikzpicture}%
\caption{Illustration of the split-aperture phased array concept for simultaneously executing three spatially tasks divided over the array.}%
\label{fig:SAPA concept}%
\end{figure}
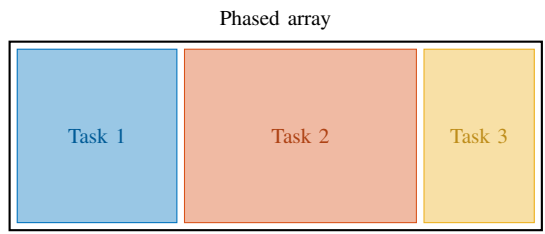

Next generation radar systems are foreseen to contain flexible apertures by further integration of digital front-end technology. The increased flexibility of these phased arrays should enable multifunction radar system to be more agile against cognitive/AI enhanced jammers or more complex operational theatres~\cite{Heras2022}. Concepts are known to increase performance of radar tasks using multiple receive channels or using multiple transmit and multiple receive channels of the array, i.e., \emph{multiple-inputs multiple-outputs} (MIMO) concepts~\cite{Melvin2013,Bergin2018}.

The flexibility of the front-end could also be leveraged by subdividing the radar tasks spatially over the array, the so-called \emph{split-aperture phased array} (SAPA) concept. An example of dividing three tasks over the array is given in Fig.~\ref{fig:SAPA concept}. In this paper, the focus is on subdividing active tracking tasks. For the tracking tasks, it will be exploited that the power-aperture budget of a radar system is usually designed for a worst-case scenario, e.g., a small, fast maneuvering target at a large distance. In case of larger targets, slower maneuvering targets, and/or targets nearer to the radar, the radar will overperform on the tracking task and the transmit time and/or track update frequency can be lowered. However, to maintain certain track requirements, the transmit time or update frequency cannot be lowered below a certain minimum value. The remaining power-aperture budget in this minimal mode may still be excessive in some cases.

Clearly, a multifunction radar system should adapt to the operational scene in which it operates. There is a vast literature on the radar resource allocation problem, e.g., see~\cite{Miranda2006,Miranda2007,Moo2015,Greco2018,Charlish2017,Charlish2020,Yan2022} to name a few. So-called bottom-up rule-based approaches are not applied in this paper, as defining the rule set may not be straightforward for this new concept. Rather, we would directly optimize tracking task quality given the limited radar resources. Hence, in this paper, we focus  on the \emph{quality of service based resource allocation model} (Q-RAM) framework for adaptive radar resource scheduling~\cite{Rajkumar1997} combined with the \emph{Van Keuk and Blackman strategy} (KBS)~\cite{VanKeuk1993} to define the performance measure of the tracking task. This combination leads to a resource allocation problem with a relatively low computational complexity compared to, e.g., neural-network, entropy, or dynamic programming based algorithms combined with other performance measures~\cite{Moo2015,Yan2022}. The Q-RAM framework relies on a conceptually simple solver to allocate the resources, but it is known to have a large computational overhead in certain cases. To alleviate the computational burden, algorithms are introduced, such as concave-majorant approximation algorithm~\cite{Hansen2004,Ghosh2006}, continuous double auction parameter selection algorithm~\cite{Charlish2015a}, or reinforcement learning~\cite{Durst2021}. The Q-RAM framework has been applied to different radar concepts including rotating radars~\cite{Yang2023}, multiple radar resources allocation problems~\cite{Irci2010}, and networks of radar systems~\cite{Nadjiasngar2015,Charlish2015}.

The goal of this work is to introduce radar resource management for the SAPA concept and demonstrate the added benefit of SAPA. In particular, the active tracking resource allocation model for SAPA will be defined based on the Q-RAM framework and by using the KBS to define the performance and resource measures. Then, the resource allocation problem is solved using the concave-majorant approximation algorithm~\cite{Hansen2004,Ghosh2006}. It will be shown that the radar system using the SAPA concept can maintain a larger number of simultaneous active tracks compared to a system using the full aperture in certain use-cases while maintaining the desired performance measure. To the authors' knowledge, this is the first work on SAPA resource management. 

The paper is organized as follows. In Sec.~\ref{sec:utility based resource management}, the utility based resource management is discussed based on the Q-RAM framework. The resource allocation model for the SAPA concept is defined in Sec.~\ref{sec:active tracking model} and solving the resource management problem is briefly discussed in Sec.~\ref{sec:solving RRM}. In Sec.~\ref{sec:example}, the effectiveness of the SAPA concept is demonstrated by a simulation example followed by the conclusions in Sec.~\ref{sec:conclusions}.

\section{Utility Based Resource Management} \label{sec:utility based resource management}

In this section, the utility based radar resource management problem is discussed. Generally speaking, efficiently allocating radar resources to different radar (tracking) tasks is performed either~\cite{Yan2022}:
\begin{enumerate*}
	\item by minimizing the system resource for given set of requirements on the different tasks, or
	\item by maximizing the combined performance of the tasks given a radar resource budget.
\end{enumerate*}
Next, the focus is on the latter set of allocation problems. The objective of the utility based radar resource management problem is to find the optimal control parameter selection $s=\{s_{t,1},…,s_{t,K} \}\in\sS$ given a set of $K$ tasks at time $t$ with an associated utility function, performance, and resource model. More specifically, the following optimization is solved~\cite{Charlish2017,Charlish2020,Yan2022}
\begin{equation}
\begin{split}
\max_{s_t} u(s_t)&=\sum_k^K\omega_ku_k\left(q_k\left(s_{t,k},e_{t,k}\right)\right), \\
\mbox{s.t.}&\sum_k^Kg_k\left(s_{t,k},e_{t,k}\right)-r_{tot}\leq 0,
\end{split}
\label{eq:constraint optimization RRM}
\end{equation}
where $q_k:\sS_k\times\sE_k\rightarrow\sQ_k$ maps the control parameter space $s_{t,k}\in\sS_k$ and the environment space $e_{t,k}\in\sE_k$ at time $t$ into the quality space for a given task $k$. The resource function of the $k$-th task is $g_k:\sS_k\times\sE_k\rightarrow\sR_k$ and the total resource is bounded by $r_{tot}\in\mathbb{R}^+$. In our case, the resource space $\sR_k$ is only the radar time budget loading. More involved resource spaces could include the transmit resources, manoeuvrings resources (organic asset), or receiving resources, see~\cite{Yan2022} for a detailed discussion. Then, $u_k:\sQ_k\rightarrow\sU=\left[0,~1\right]$ denotes the utility function of the $k$-th task with task weight $\omega_k\in\left[0,~1\right]$ normalized to $\sum \omega_k =1$.

Formulation~\eqref{eq:constraint optimization RRM} allows to allocate resources in an adaptive manner by obtaining a solution for a certain time window and solve it repetitively in time to adjust to the current environment and/or mission specific requirements. Mission specific requirements are included by, e.g., selection of weights $\omega_k$, type of utility functions $u_k$, etc. However, systematic translation of these requirements into~\eqref{eq:constraint optimization RRM} is still an open research topic and it requires deep expert knowledge.

In Sec.~\ref{sec:active tracking model}, the functions and spaces in~\eqref{eq:constraint optimization RRM} for the active tracking tasks are discussed and solving~\eqref{eq:constraint optimization RRM} is briefly considered in Sec.~\ref{sec:solving RRM}. For notational simplicity, the input arguments to the quality function $q_k$, the resource function $g_k$, and the utility function $u_k$ will be omitted in subsequent sections.

\section{Active Tracking Resource Allocation Model} \label{sec:active tracking model}

The resource allocation model for the active tracking tasks are defined for the SAPA concept in this section. The model is based on the Van Keuk and Blackman strategy~\cite{VanKeuk1993} that is originally used to adaptively select the revisit interval for a tracking task. The KBS is an empirically determined function to maintain a certain maximum major axis of the posterior track covariance matrix, i.e., angular estimation error in u-v space. The strategy is based on assuming a Swerling I target amplitude fluctuation model with well-separated point targets in space. The strategy takes into account a complete tracking loop, including beam position over time, data association, and tracker dynamics based on a Singer model.

As shown next, the KBS can be used in the resource allocation problem~\eqref{eq:constraint optimization RRM}. It is assumed that the tracking tasks can only be horizontally distributed over the phased array.

\subsection{Control Space}

The control parameters space at time $t$ for the $k$-th task is defined by the coherent integration time $T_{d,k}\in\mathbb{R}^+$, the update frequency $f_{t,k}\in\mathbb{R}^+$, and the number of elements $N_{h,k}\in\mathbb{N}$ used in the horizontal direction, denoted as $s_{t,k}=\{T_{d,k}, f_{t,k}, N_{h,t}\}$. Note that, the parameters defining $s_{t,k}$ can be different per active tacking task.

\subsection{Environmental Space}

The environmental space $e_{t,k}\!=\!\{R_{t,k},\!\theta_{t,k},\!\sigma_{t,k},\!\Theta_{t,k},\!\Sigma_{t,k}\}$ for the active tracking model based on the KBS strategy at time $t$ for the $k$-th target is defined by the target range $R_{t,k}\in\mathbb{R}^+$, the bearing $\theta_{t,k}\in\mathbb{R}$, the radar cross section $\sigma_{t,k}\in\mathbb{R}^+$, and the standard deviation and time correlation $(\Theta_{t,k}, \Sigma_{t,k})\in\mathbb{R}^+\times\mathbb{R}^+$ of Singer movement model. Standard tracking techniques can estimate the parameters in the environmental space $e_{t,k}$.

\subsection{Quality Function}

For the $k$-th active tracking task, the parameter of interest is the angular estimation error and it is given by~\cite{VanKeuk1993}
\begin{equation}
q_k = \theta_{bw}v_0,
\label{eq:quality active track}
\end{equation}
where $v_0\in\mathbb{R}^+$ denotes the track-sharpness, $\theta_{bw}=\frac{\theta_{bw,0}}{\theta_{t,k}}$ is the broadened half beamwidth where the half beamwidth at boresight is taken as $\theta_{bw,0}=\frac{\alpha_{bw}}{N_{h,k}}$ assuming half wavelength element spacing with beamwidth factor $\alpha_{bw}\in\mathbb{R}^+$. The track-sharpness $v_0$ is obtained by finding the root of~\cite{Charlish2015}
\begin{equation}
	1+\left(\frac{\beta}{2}+2\right)v_0^2-\alpha\beta v_0^{2.4}=0,
	\label{eq:track sharpness}
\end{equation}
with
\begin{subequations}
\begin{align}
	\alpha &= 0.4f_{t,k} \left( \frac{R_{t,k}\theta_{bw}\sqrt{\Sigma_{t,k}}}{\Theta_{t,k}} \right)^{0.4}, \label{eq:alpha} \\
	\beta  &= \xi \SNn - \ln P_{fa} \label{eq:beta},
\end{align}
\end{subequations}
where $\xi\in(0,~1]$ denotes a cross-talk loss function, $\SNn\in\mathbb{R}^+$ is the \emph{signal-to-noise ratio} (SNR) without angular pointing error, and $P_{fa}\in\mathbb{R}^+$ defines the false alarm rate. Note that, in contrast to~\cite{Charlish2015,Charlish2017}, the angular estimation error and not the track-sharpness is used as a quality function in~\eqref{eq:quality active track}, because we should take into account that the beamwidth varies when changing the number of horizontal elements $N_{h,k}$ in the control space. 

It is assumed that spatially divided tasks over the array will come at the cost of cross-talk between tasks. Therefore, the following loss function is applied
\begin{equation}
	\xi = 0.8+0.2\frac{N_{h,k}}{N_{hT}},
	\label{eq:cross-talk loss}
\end{equation}
where $N_{hT}\in\mathbb{N}^+$ denotes the total number of antenna elements in the horizontal dimension. The $\SNn$ in~\eqref{eq:beta} is given by~\cite{Richards2010}\footnote{The antenna gain is $G\approx\frac{4\pi\eta_\alpha A \cos\left(\theta_{t,k}\right)}{\lambda^2}$~\cite{Richards2010} and the array area is $A=\frac{\lambda}{2}N_{h,k}\frac{\lambda}{2}N_{vT}$, which leads to $G\approx\pi\eta_\alpha N_{h,k}N_{vT}\cos\left(\theta_{t,k}\right)$.}
\begin{equation}
	\SNn = k_{rad} \frac{N^3_{h,k}T_{d,k}\cos^2\left(\theta_{t,k}\right)\sigma_{t,k} }{R_{t,k}^4},
	\label{eq:SN0}
\end{equation}
where the constant $k_{rad}\in\mathbb{R}^+$ is
\begin{equation}
	k_{rad}=\frac{P_{avg}\lambda^2\eta_\alpha^2 N_{vT}^2}{64\pi N_{hT} k_bT_0FL_s},
	\label{eq:}
\end{equation}
with $P_{avg}\in\mathbb{R}^+$ denoting the average transmit power when all elements are active, $\lambda\in\mathbb{R}^+$ is the wavelength, $\eta_\alpha\in\mathbb{R}^+$ is the aperture efficiency, $N_{vT}\in\mathbb{N}^+$ defining the total number of antenna elements in the vertical dimension, $k_b\in\mathbb{R}^+$ is the Boltzmann’s constant, $T_0\in\mathbb{R}^+$ is the receive system noise temperature, $F\in\mathbb{R}^+$ is the noise floor, and $L_s\in\mathbb{R}^+$ denoting the system losses.

In our paper, the maximum value of $\SNn$ is limited to 40\,dB and the minimum value to 10\,dB. Due to practical limitations of a radar system such as calibration errors, phase errors, etc., we assume that the $\SNn$ cannot exceed a measurement accuracy corresponding to $\SNn=40$\,dB and $v_0$ in~\eqref{eq:quality active track} is lower bounded using $\SNn=40$\,dB. If $\SNn$ is below 10\,dB, it will imply that the target will not be detected and, hence, the quality $q_k$ cannot be a number.

\subsection{Resource Function}

The expected steady-state resource on the radar system is approximated by~\cite{Charlish2015}\footnote{Compare to~\cite{Charlish2015}, we have added the split aperture ratio $\frac{N_{h,k}}{N_{hT}}$ to~\eqref{eq:resource function}.}
\begin{equation}
	g_k = n_l T_{d,k}f_{t,k}\frac{N_{h,k}}{N_{hT}},
	\label{eq:resource function}
\end{equation}
where $n_l\in\mathbb{R}^+$ defines the expected number of looks given by~\cite{VanKeuk1993}
\begin{subequations}
\begin{equation}
	n_l=\frac{1}{P_D} \left(1+\left(\gamma v_0^2\right)^2 \right)^{1/2},
	\label{eq:nr looks}
\end{equation}
with $\gamma\in\mathbb{R}^+$ and the probability of detection $P_D\in\mathbb{R}^+$, assuming a Swerling I target fluctuation model, are
\begin{align}
\gamma &\cong 1+14\left(\frac{\vert \ln P_{fa}\vert}{\xi\SNn}\right)^{1/2}, \label{eq:nr looks dist} \\
P_D 	 &= P_{fa}^{\frac{1}{1+\xi\SNn}}. \label{eq:nr looks Pd} 
\end{align}
\end{subequations}
Equivalent to the quality function $q_k$ in \eqref{eq:quality active track}, the $\SNn$ is capped at 40\,dB and if the $\SNn$ is below 10\,dB, then the resource $g_k$ will not be a number. The resource function $g_k$ in~\eqref{eq:resource function} also accounts for the potential need of multiple observations of the target due to a low detection probability.

Note that the number of elements $N_{h,k}$ that are selected have an impact on both the quality $q_k$ in \eqref{eq:quality active track} and on the resource function $g_k$ in \eqref{eq:resource function}. The number of elements influences, amongst others, the beamwidth $\theta_{bw}$, the $\SNn$ via the transmit power and antenna gain, and via the cross-talk loss function.

\subsection{Utility Function}

The utility function $u_k$ characterizes the level of satisfaction of each task. Selecting a suitable function depends on the system specifications, mission statements, and expert knowledge. In this paper, a linear utility function is applied given by
\begin{equation}
	u_k = \max\left(\min\left(\frac{q_k-q_{k,min}}{q_{k,max}-q_{k,min}},1\right), 0\right),
	\label{eq:linear utility}
\end{equation}
where $q_{k,min},q_{k,max}\in\mathbb{R}$ are the minimum and maximum values for the angular estimation error for the $k$-th task. The $\min$ and $\max$ functions in the utility function~\eqref{eq:linear utility} are used to bound the angular estimation error on the interval $[0,~1]$. Note that, for the track quality $q_k$, a smaller angular estimation error implies an improved track accuracy.

\section{Solving SAPA Radar Resource Management} \label{sec:solving RRM}

Solving the optimization~\eqref{eq:constraint optimization RRM} can be performed by using different Q-RAM solvers~\cite{Rajkumar1997}, including convex hull approximation algorithm~\cite{Hansen2004,Ghosh2006}, continuous double auction parameter selection algorithm~\cite{Charlish2015a}, and reinforcement learning~\cite{Durst2021}.

The Q-RAM formulation is well-known to solve the Karush-Kuhn-Tucker optimality conditions for a discrete control space $s_{t,k}$ under the assumptions that the utility $u_k$ and resource $g_k$ functions are concave functions in $s_{t,k}$ given $e_{t,k}$~\cite{Irci2010,Charlish2017}. A basic algorithm to solve~\eqref{eq:constraint optimization RRM} will iteratively allocates resource increments to tasks, starting from zero resource, to maximize the combined utility. This basic algorithm consists of the following steps~\cite{Rajkumar1997}:
\begin{enumerate}
	\item Generate the utility $u_k$ and resource $g_k$ set-points for all discrete control parameter in $\sS$.
	\item Construct the concave-majorant for each task $k$ by a convex hull operation.
	\item Order the set-points of the concave-majorant for all tasks in descending order based on the marginal utility. The marginal utility is the difference in utility divided by the difference in resource between set-points.
	\item Traverse over the sorted list with highest marginal utility to allocate resource until no resource remains.
\end{enumerate}

This algorithm has certain drawbacks. The computational complexity can be high when the dimension of the discrete control parameter space is large. Traversal techniques to approximate the concave-majorant can reduced the overhead and, in Sec.~\ref{sec:example}, the first-order fast traversal is applied~\cite{Hansen2004,Ghosh2006}. Also, the algorithm to solve Q-RAM achieves near-optimal solutions due to sub-optimal stopping conditions, as the optimal solution might not lie on the concave-majorant~\cite{Irci2010,Charlish2017}. To minimize sub-optimality, the discrete control space should contain sufficient granularity.

\section{Split-Aperture Resource Allocation Example} \label{sec:example}

In this section, the split-aperture resource allocation problem of the SAPA concept is compared to the resource allocation where the radar cannot split the aperture to demonstrate the benefits of the SAPA concept. In Sec.~\ref{subsec:simulation setting}, the simulation setting is provided and the results are discussed in Sec.~\ref{subsec:Simulation results}.

\subsection{Simulation Setting} \label{subsec:simulation setting}

\begin{table}[!t]
\caption{The parameter ranges of the Singer movement model for different target types~\cite{Charlish2015a}.}\label{tbl:Singer Model}
\centering
\begin{tabular}{|c|c|c|} 
 \hline
  & $\Theta_{t,k}~[m/s^2]$ & $\Sigma_{t,k}~[s]$  \\
 \hline\hline
 Type I & 20-35 & 10-20 \\ \hline
 Type II & 0-5 & 1-4 \\ \hline
\end{tabular}
\end{table}

The simulation setting is discussed in this section, which includes the applied scene, the target parameters, the radar parameters, and the resource allocation parameters. The scene will contain 200 targets where the environmental space $e_{t,k}$ of the $k$-th target is defined as follows. The bearing is a realization from a uniform distribution on $\theta_{t,k}\sim\mathpzc{U}(-60, 60)^{\circ}$ and the RCS on $\sigma_{t,k}\sim\mathpzc{U}(-10, 10)\,\text{dBm}^2$. The targets have an equal chance in being type I or type II and, after the type is randomly selected, the standard deviation $\Theta_{t,k}$ and time correlation $\Sigma_{t,k}$ are randomly drawn from a uniform distribution with the domain as indicated in Tab.~\ref{tbl:Singer Model}. Two scenes will be simulated. In the first scene, the targets are placed randomly in range according to $R_{t,k}\sim\mathpzc{U}(10, 70)$\,km and, in the second scene, according to $R_{t,k}\sim\mathpzc{U}(10, 250)$\,km. The task weights are uniformly drawn from $\omega_k~\sim\mathpzc{U}(0.2, 0.8)$ and then normalized.

For the radar, the following constant is used $k_{rad}=2.662\cdot 10^{21}\,m^2/s$, the false alarm rate is set at $P_{fa}=10^{-4}$, and the total amount of elements in the horizontal dimension is $N_{hT}=48$.

For the control space of each tracking task, the coherent integration time $T_{d,k}$ can be chosen from the discrete set $[4,~4.6,\ldots,64]$\,ms, the update rate $f_{t,k}$ from $[0.1,~0.2,\ldots,6]$\,Hz, and the number of horizontal elements $N_{h,k}$ from $[6, 12,\ldots,48]$. Note that the split-aperture allocation problem also includes $N_{h,k}=48$ elements in the control space, i.e., it can assign the full array for a certain task if necessary. To compare, the resource allocation is also solved for the full-aperture case for which the control space for the number of horizontal elements is $48$. The utility function has as its minimal value $q_{k,min}=3$\,mrad and its maximum value $q_{k,max}=1$\,mrad, which implies this $u_k=1$ if $q_k\leq 1$\,mrad and $u_k=0$ if $q_k\geq3$\,mrad. For the simulation example, $N_{MC}=100$ Monte Carlo runs are conducted where each run has a new realization of the environmental space $e_{t,k}$ and weights $\omega_k$.

\begin{figure}[!t]%
\centering%
\input{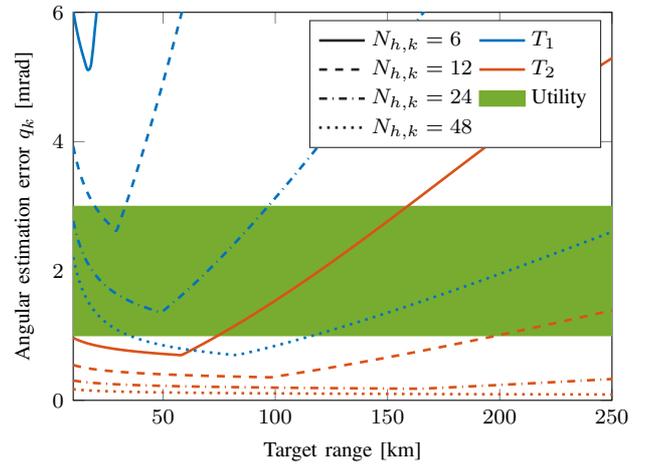}%
\caption{The quality function of the active track task $q_k$. The angular estimation error is displayed for different target ranges, number of horizontal elements $N_{h,k}$, and target parameters $T_1$,$T_2$ for an update rate of $f_{t,k}=1$\,Hz.}%
\label{fig:quality function}%
\end{figure}

Fig.~\ref{fig:quality function} shows the quality function $q_k$ of the $k$-th active tracking task, i.e., the angular estimation error. The figure shows target $T_1$ with parameters $\sigma_{t,k}=0.1\,\text{m}^2$, $T_{d,k}=64$\,ms, $\Sigma_{t,k}=35\,\text{m/s}^2$, $\Theta_{t,k}=10$\,s, $\theta_{t,k}=60^{\circ}$ and target $T_2$ with parameters $\sigma_{t,k}=10\,\text{m}^2$, $T_{d,k}=20$\,ms, $\Sigma_{t,k}=0.1\,\text{m/s}^2$, $\Theta_{t,k}=4$\,s, $\theta_{t,k}=0^{\circ}$. Target $T_1$ represents the worst-case target parameters and $T_2$ is an optimistic case in terms of the radar resources. The green area in Fig.~\ref{fig:quality function} indicates the area between $q_{k,min}$ and $q_{k,max}$ of the utility function. If the angular estimation error is smaller than $1$\,mrad then the radar would use too many resources. For angular estimation error bigger than $3$\,mrad, the allocation problem would use too little resources. Clearly, at a closer range or for different target parameters, the radar can utilize the split-aperture for resource allocation. As discussed before, it is assumed that the transmit time or update frequency cannot be lowered below a certain minimum value to maintain certain track requirements such as Doppler resolution. Note that, at close range, there is a discontinuity in the lines caused by our assumption that the SNR $\SNn$ cannot exceed 40\,dB.

\subsection{Simulation Results} \label{subsec:Simulation results}

\begin{figure}%
\centering%
\input{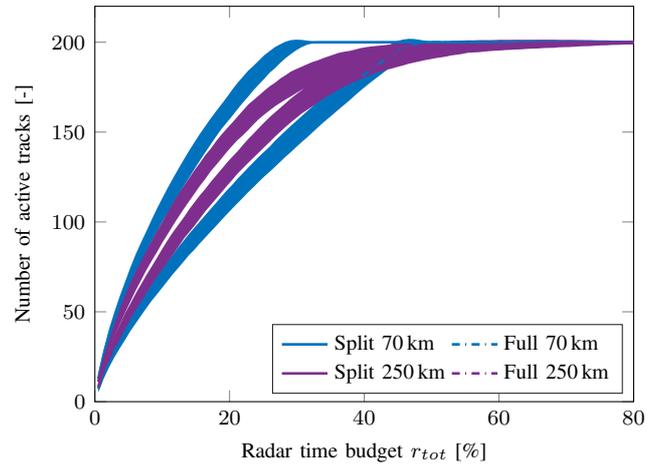}%
\caption{The number of active tracks given a radar time budget for the split-aperture and full-aperture resource problem in a scene with targets at a maximum range of 70\,km or 250\,km. The curve is the mean over $N_{MC}=100$ Monte Carlo runs and the area indicates $\pm 2\sigma$ deviation.}%
\label{fig:active tracks}%
\end{figure}

The simulation results of using the split-aperture resource allocation with respect to using the full-aperture resource allocation is discussed next. Fig.~\ref{fig:active tracks} shows the number of active tracks for a certain radar time budget, i.e., the number of active tracks for a given maximum resource $r_{tot}$, for a scene with targets at maximum range of 70\,km or 250\,km. The number of active tracks imply the number of tracking tasks that the resource allocation solver assigns a resource larger than zero. The curve is the mean over $N_{MC}=100$ Monte Carlo runs and the area indicates $\pm 2\sigma$ deviation. The figure clearly indicates that the split-aperture resource allocation problem has a higher number of active tracks compared to the full-aperture resource allocation problem in both scenes. Moreover, to track all 200 targets, the split-aperture resource allocation problem requires significantly less resources from the radar system. The difference between the split-aperture and full-aperture is smaller in case that the maximum target range is at 250\,km, because the full array is in both allocation problems necessary to observe small targets at a large distance with sufficient quality. 

The total utility $u$ for a certain radar time budget $r_{tot}$ for a scene with targets at maximum range of 70\,km or 250\,km is given in Fig.~\ref{fig:total utility}. In line with the discussion of Fig.~\ref{fig:active tracks}, the split-aperture resource allocation problem achieves an increased total utility compare to the full-aperture resource allocation problem in both scenes. The total utility monotonically increases to 1 as expected.

\begin{figure}%
\centering%
\input{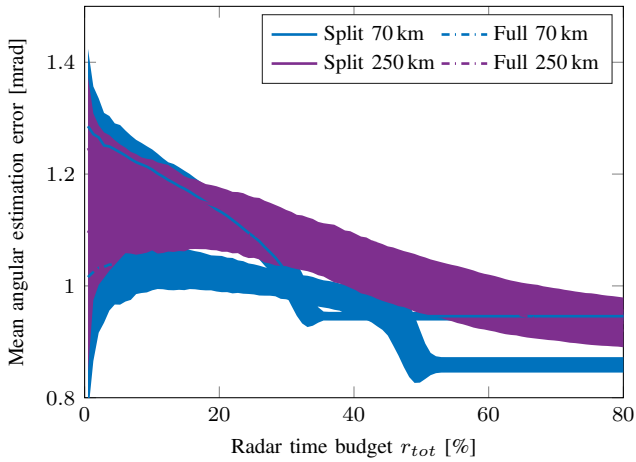}%
\caption{The mean angular estimation error given a radar time budget for the split-aperture and full-aperture resource problem in a scene with targets at a maximum range of 70\,km or 250\,km. The curve is the mean over $N_{MC}=100$ Monte Carlo runs and the area indicates $\pm 2\sigma$ deviation.}%
\label{fig:track sharpness}%
\end{figure}

Fig.~\ref{fig:track sharpness} provides the mean angular estimation error for a certain radar time budget $r_{tot}$. The mean value is computed from all tasks that have a non-zero resource assigned to it. The figure highlights if the resource allocation problem only assigns resources such that $u_k=1$ ($q_k\leq1$\,mrad) or if it chooses to distribute resources among tasks with a slight but acceptable decrease in performance $0<u_k<1$ ($1>q_k>3$\,mrad). If also considering Fig.~\ref{fig:active tracks} and Fig.~\ref{fig:track sharpness}, then it can be concluded that the angular estimation error is traded-off to achieve a higher number of active tracks in all cases, as is the expectation of this type of resource allocation problems. In case when sufficient radar time budget $r_{tot}$ is available, then the total utility should be one and the mean angular estimation error should equal $1$\,mrad. If it is below $1$\,mrad, then it indicates that the resource allocation assigns a bit more resources than strictly necessary. In such a case, either the control space $\mathcal{S}_k$ has insufficient granularity or the minimum in the control space $s_{t,k}$ still results in an excessive aperture-time budget for that task given the environmental space $e_{t,k}$. In Fig.~\ref{fig:track sharpness}, the mean angular estimation error descends furthest below the $1$\,mrad for the full-aperture in the scene with targets at maximum of 70\,km compared to the other cases implying that excessive aperture-time budget is allocated.

\begin{figure}%
\centering%
%
\definecolor{mycolor1}{rgb}{0.00000,0.44700,0.74100}%
\definecolor{mycolor2}{rgb}{0.85000,0.32500,0.09800}%
\definecolor{mycolor3}{rgb}{0.92900,0.69400,0.12500}%
\definecolor{mycolor4}{rgb}{0.49400,0.18400,0.55600}%
\begin{tikzpicture}[font=\footnotesize]

\begin{axis}[%
width=0.8\columnwidth,
height=0.586\columnwidth,
at={(0\columnwidth,0\columnwidth)},
scale only axis,
ybar=0, 
bar width=4pt,
tick align=inside,
xlabel={Number of horizontal elements $N_{h,k}$ [-]},
ymin=0,
ymax=55,
ylabel={Number of tasks [-]},
symbolic x coords={6,12,18,24,30,36,42,48},
xtick = data,
scaled y ticks = false,
]

\addplot[style={mycolor1,fill=mycolor1,mark=none}]
            coordinates {(6, 1) (12, 20) (18,37) (24,27) (30,12) (36,6) (42,3) (48,2)};

\addplot[style={mycolor2,fill=mycolor2,mark=none}]
            coordinates {(6, 2) (12, 28) (18,49) (24,42) (30,23) (36,12) (42,6) (48,4)};
						
\addplot[style={mycolor3,fill=mycolor3,mark=none}]
            coordinates {(6, 2) (12, 31) (18,54) (24,49) (30,30) (36,16) (42,9) (48,8)};

\addplot[style={mycolor4,fill=mycolor4,mark=none}]
            coordinates {(6, 1) (12, 31) (18,54) (24,49) (30,29) (36,16) (42,8) (48,12)};

\legend{$r_{tot}=10$\%,$r_{tot}=20$\%,$r_{tot}=30$\%,$r_{tot}=40$\%}

%
%

\end{axis}
\end{tikzpicture}%
\caption{The mean number of tasks with the assigned number of horizontal elements used for varying radar time budgets $r_{tot}$ in a scene with targets at a maximum range of 70\,km. The curve is the mean over $N_{MC}=100$ Monte Carlo runs.}%
\label{fig:tracks vs elements}%
\end{figure}
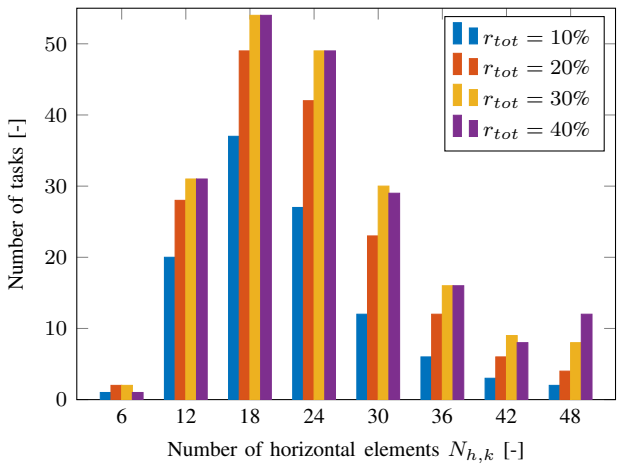

The mean number of active tracks with the assigned number of horizontal elements $N_{h,k}$ is given in Fig.~\ref{fig:tracks vs elements} in a scene with targets at a maximum range of 70\,km. For the radar time budgets $r_{tot}=\{10,20,30\}\%$, the total number of active tracks with a resource assigned is smaller than the total amount of 200 in the $r_{tot}=40\%$ case. For the $r_{tot}=30\%$ case, the mean number of active tasks is slightly below 200 and the total utility is just below $u=1$ see Fig.~\ref{fig:active tracks} and Fig.~\ref{fig:track sharpness}. Hence, the $r_{tot}=30\%$ case has more tasks with a smaller number of horizontal elements assigned than for the $r_{tot}=40\%$ case in Fig.~\ref{fig:tracks vs elements}.

To summarize, for active tracking of many targets, the split-aperture phased array resource allocation concept can lead to a more efficient utilization of the radar resources and an increased number of targets in active track when the radar has limited resources.

\begin{figure}%
\centering%
\input{total_utility.tex}%
\caption{The total utility given a radar time budget for the split-aperture and full-aperture resource problem in a scene with targets at a maximum range of 70\,km or 250\,km. The curve is the mean over $N_{MC}=100$ Monte Carlo runs and the area indicates $\pm 2\sigma$ deviation.}%
\label{fig:total utility}%
\end{figure}

\section{Conclusions} \label{sec:conclusions}

It is foreseen that the front-end technology of phased arrays becomes more flexible in the next generation of radar systems and it could potentially support the split-aperture phased array concept. To demonstrate the potential advantages of this flexibility, in this paper, the SAPA radar resource management has been solved for loading the radar with many active tracking tasks. In particular, the Van Keuk and Blackman strategy has been applied for the utility based resource management problem. To the authors’ knowledge, this is the first work on SAPA radar resource management in the literature. The simulation study showed that the SAPA concept can increase the number of active tracks compared to an array without SAPA concept while maintaining the necessary track quality for the same radar time budget.

For future research, the SAPA resource management can be extended to also include the vertical dimension of the array and to included tracker independent performance measures similar to the (Bayesian) Cram\'{e}r-Rao lower bound. Moreover, the tracking tasks can be considered blocks in the number of horizontal elements, the number of vertical elements, and time which should be scheduled on the array over time where these tasks cannot intersect. This set of constrains is not taken into account in this paper and it is a topic for future research.

\section*{Acknowledgment}
This project has received funding from the European Union’s Preparatory Action on Defence Research under grant agreement No. 882407.

\bibliographystyle{IEEEtran}

\bibliography{bibliography}

\begin{thebibliography}{10}
\providecommand{\url}[1]{#1}
\csname url@samestyle\endcsname
\providecommand{\newblock}{\relax}
\providecommand{\bibinfo}[2]{#2}
\providecommand{\BIBentrySTDinterwordspacing}{\spaceskip=0pt\relax}
\providecommand{\BIBentryALTinterwordstretchfactor}{4}
\providecommand{\BIBentryALTinterwordspacing}{\spaceskip=\fontdimen2\font plus
\BIBentryALTinterwordstretchfactor\fontdimen3\font minus
  \fontdimen4\font\relax}
\providecommand{\BIBforeignlanguage}[2]{{%
\expandafter\ifx\csname l@#1\endcsname\relax
\typeout{** WARNING: IEEEtran.bst: No hyphenation pattern has been}%
\typeout{** loaded for the language `#1'. Using the pattern for}%
\typeout{** the default language instead.}%
\else
\language=\csname l@#1\endcsname
\fi
#2}}
\providecommand{\BIBdecl}{\relax}
\BIBdecl

\bibitem{Heras2022}
M.~L.~G. Heras, A.~C. Garcia, J.~C. Soriano, J.~L.~G. {De La Haba}, I.~{Le
  Roy-Naneix}, S.~Kemkemian, M.~Thorsell, M.~Brandfass, P.~Brouard, T.~Boman,
  S.~Durst, A.~Nanni, J.~J.~M. {De Wit}, U.~Calfa, and M.~Sakalas, ``{CROWN}
  project, towards a {E}uropean multifunction {AESA} system,'' in \emph{Proc.
  of the IEEE Int. Symp. on Phased Array Systems and Technology}, Waltham, MA,
  USA, Dec. 2022, pp. 1--8.

\bibitem{Melvin2013}
W.~L. Melvin and J.~A. Scheer, Eds., \emph{Principles of modern radar: advanced
  techniques}.\hskip 1em plus 0.5em minus 0.4em\relax SciTech, 2013.

\bibitem{Bergin2018}
J.~Bergin and J.~H. Guerci, \emph{{MIMO} Radar: theory and application}.\hskip
  1em plus 0.5em minus 0.4em\relax Artech House, 2018.

\bibitem{Miranda2006}
S.~L.~C. Miranda, C.~J. Baker, K.~Woodbridge, and H.~D. Griffiths,
  ``Knowledge-based resource management for multifunction radar: a look at
  scheduling and task prioritization,'' \emph{IEEE Signal Processing Magazine},
  vol.~23, no.~1, pp. 66--76, 2006.

\bibitem{Miranda2007}
------, ``Comparison of scheduling algorithms for multifunction radar,''
  \emph{IET Radar, Sonar and Navigation}, vol.~1, no.~6, pp. 414--424, 2007.

\bibitem{Moo2015}
P.~W. Moo and Z.~Ding, \emph{Adaptive radar resource management}.\hskip 1em
  plus 0.5em minus 0.4em\relax Elsevier, 2015.

\bibitem{Greco2018}
M.~S. Greco, F.~Gini, P.~Stinco, and K.~Bell, ``Cognitive radars: On the road
  to reality: Progress thus far and possibilities for the future,'' \emph{IEEE
  Signal Processing Magazine}, vol.~35, no.~4, pp. 112--125, 2018.

\bibitem{Charlish2017}
A.~Charlish and F.~Hoffmann, ``Cognitive radar management,'' in \emph{Novel
  radar techniques and applications: Volume 2: Waveform diversity and cognitive
  radar, and target tracking and data fusion}.\hskip 1em plus 0.5em minus
  0.4em\relax Institution of Engineering and Technology, 2017, ch.~3, pp.
  157--193.

\bibitem{Charlish2020}
A.~Charlish, F.~Hoffmann, C.~Degen, and I.~Schlangen, ``The development from
  adaptive to cognitive radar resource management,'' \emph{IEEE Aerospace and
  Electronic Systems Magazine}, vol.~35, no.~6, pp. 8--19, 2020.

\bibitem{Yan2022}
J.~Yan, H.~Jiao, W.~Pu, C.~Shi, J.~Dai, and H.~Liu, ``Radar sensor network
  resource allocation for fused target tracking: A brief review,''
  \emph{Information Fusion}, vol. 86-87, no. June, pp. 104--115, 2022.

\bibitem{Rajkumar1997}
R.~Rajkumar, C.~Lee, J.~Lehoczky, and D.~Siewiorek, ``Resource allocation model
  for {QoS} management,'' in \emph{Proc. of the Real-Time Systems Symp.}, San
  Francisco, CA, USA, Dec. 1997, pp. 298--307.

\bibitem{VanKeuk1993}
G.~{Van Keuk} and S.~S. Blackman, ``On phased-array radar tracking and
  parameter control,'' \emph{IEEE Trans. on Aerospace and Electronic Systems},
  vol.~29, no.~1, pp. 186--194, 1993.

\bibitem{Hansen2004}
J.~P. Hansen, S.~Ghosh, R.~Rajkumar, and J.~Lehoczky, ``Resource management of
  highly configurable tasks,'' in \emph{Proc. of the Int. Parallel and
  Distributed Processing Symp.}, Santa Fe, NM, USA, Apr. 2004, pp. 1615--1622.

\bibitem{Ghosh2006}
S.~Ghosh, R.~Rajkumar, J.~Hansen, and J.~Lehoczky, ``Integrated {QoS}-aware
  resource management and scheduling with multi-resource constraints,''
  \emph{Real-Time Systems}, vol.~33, pp. 7--46, 2006.

\bibitem{Charlish2015a}
A.~Charlish, K.~Woodbridge, and H.~Griffiths, ``Phased array radar resource
  management using continuous double auction,'' \emph{IEEE Trans. on Aerospace
  and Electronic Systems}, vol.~51, no.~3, pp. 2212--2224, 2015.

\bibitem{Durst2021}
S.~Durst and S.~Br\"{u}ggenwirth, ``Quality of service based radar resource
  management using deep reinforcement learning,'' in \emph{Proc. of the IEEE
  Radar Conf.}, Atlanta, GA, USA, May 2021.

\bibitem{Yang2023}
N.~Yang, Y.~Feng, and J.~Yu, ``Quality of service based resource management for
  rotating phased array radar,'' in \emph{Proc. of the 5th Int. Conf. on
  Electronic Engineering and Informatics}, Wuhan, China, Jun. 2023, pp.
  385--390.

\bibitem{Irci2010}
A.~Irci, A.~Saranli, and B.~Baykal, ``Study on {Q-RAM} and feasible directions
  based methods for resource management in phased array radar systems,''
  \emph{IEEE Trans. on Aerospace and Electronic Systems}, vol.~46, no.~4, pp.
  1848--1864, 2010.

\bibitem{Nadjiasngar2015}
R.~Nadjiasngar and A.~Charlish, ``Quality of service resource management for a
  radar network,'' in \emph{Proc. of the IEEE Radar Conf.}, Johannesburg, South
  Africa, Oct. 2015, pp. 344--349.

\bibitem{Charlish2015}
A.~Charlish and R.~Nadjiasngar, ``Quality of service management for a
  multi-mission radar network,'' in \emph{Proc. of the IEEE 6th Int. Workshop
  on Computational Advances in Multi-Sensor Adaptive Processing}, Cancun,
  Mexico, Dec. 2015, pp. 289--292.

\bibitem{Richards2010}
M.~A. Richards, J.~A. Scheer, and W.~A. Holm, \emph{Principles of modern radar:
  basic principles}.\hskip 1em plus 0.5em minus 0.4em\relax Institution of
  Engineering and Technology, 2010.

\end{thebibliography}

\end{document}